# Wikidata Vandalism Detection
## The Loganberry Vandalism Detector at WSDM Cup 2017

Qi Zhu
qiz3@illinois.edu

Hongwei Ng
hongwei2@illinois.edu

Liyuan Liu
ll2@illinois.edu

Ziwei Ji
ziweiji2@illinois.edu

Bingjie Jiang
bingjie3@illinois.edu

Jiaming Shen
js2@illinois.edu

Huan Gui
huangui2@illinois.edu

## ABSTRACT
Wikidata is the new, large-scale knowledge base of the Wikimedia Foundation. As it can be edited by anyone, entries frequently get vandalized, leading to the possibility that it might spread of falsified information if such posts are not detected. The WSDM 2017 Wiki Vandalism Detection Challenge requires us to solve this problem by computing a vandalism score denoting the likelihood that a revision corresponds to an act of vandalism and performance is measured using the ROC-AUC obtained on a held-out test set. This paper provides the details of our submission that obtained an ROC-AUC score of 0.91976 in the final evaluation.

## Keywords
Wikidata vandalism detection

## 1. IMPLEMENTATION DETAILS

In this section, we provide the details of our system design, our data processing steps, and the features and learning algorithms that make up our model.

### 1.1 Data sampling

The training data provided by the organizers consists of 65009960 revisions, of which 176320 were positive examples (revisions corresponding to vandalized posts) and the rest negative ones. We kept 174427 of the positive examples and discarded the rest as they were not usable due to missing values in the revision (see section 1.2).

As the training dataset is too big for our system to handle, we decided to keep all the positive examples and sample without replacement 2.5 times the number of positive examples from the negative examples. The number of negative examples sampled from each of the 21 batches of revision data provided by the organizers is in proportion to their size so that more negative examples are sampled from batches with more examples. Another reason we sub-sampled the negative examples is to ensure that our model do not overfit to the large number of negative examples. Having reduced the size of our training set to 174427 positive and 4360675 negative examples, we further sampled 80% of the dataset for training and 20% for validation.

### 1.2 Features

The features we use are built on top of the features described in [5] Specifically, we generated features using the code provided by organizers and processed them.

Features that were missing from more than 25% of the training data were excluded and we imputed the remaining ones with missing values using the corresponding median value for each of those features in the training data. Note that these imputed values computed from the training data are used later for imputing missing features in the validation and test data. In particular, we discarded the revision ID, user ID, user name, group ID, timestamp from the 47 features generated from the code provided by [4] as we feel that these are potentially misleading features. We also excluded 7 geographical features as we were not able to get the code from [4] to generate them.

Lastly, we added spam features. From our observation, categorical features like userName/groupID have a large data range and they are not meaningful features as we investigated in feature importance study. So we introduced spam counting feature, which counted number of spam and number of occurrence in training dataset. Instead using discrete features, **e.g.** {'UserID':1..1000}, we used spam probability and spam counts for specific categorical value, **e.g.** {'User1': 0.85,300}, which means revisions of user1 are reported as spam for 300 times and the spam ratio is 85%. These features effectively captured quality of user/item in our experiment.

### 1.3 Learning algorithms

Due to the large size of the dataset, we only considered learning algorithms that can scale well. Hence, we experimented with the Logistic Regression trained with stochastic gradient descent [1] and Extremely Randomized Trees [3] implemented in Scikit-Learn library [6] and XGBoost [2].

## 2. EVALUATION

We trained the algorithms listed in section 1.3 and selected the model that maximized the ROC-AUC score for each of them using 5 folds cross validation maximizing ROC-AUC. Finally we evaluated the performance of each model on the validation set. The result is shown in Table 1 and the ROC curves and precision recall curves are shown in Figure 1 and 2 respectively. Based on the result shown in Table 1, we picked the XGBoost model as the final model. The performance of this model on the final test data released by the organizers is listed in Table 2.

| Metric | ROC | ACC | P | R | F |
|---|---|---|---|---|---|
| LR | 0.9382 | 0.9667 | 0.04 | 0.82 | 0.07 |
| ET | 0.9808 | 0.9903 | 0.12 | 0.83 | 0.20 |
| XGB | **0.9868** | 0.9880 | 0.10 | 0.87 | 0.18 |

Table 1: Evaluation of our models on validation set. Logistic Regression (LR), Extremely Randomized Trees (ET), and XGBoost (XGB).

| Metric | ROC | PR | Acc | P | R | F |
|---|---|---|---|---|---|---|
| XGB | 0.920 | 0.337 | 0.929 | 0.011 | 0.767 | 0.022 |

Table 2: Evaluation of our model on test set.

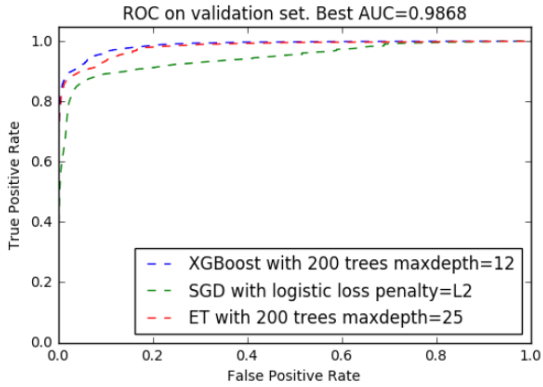

Figure 1: ROC curves for our models

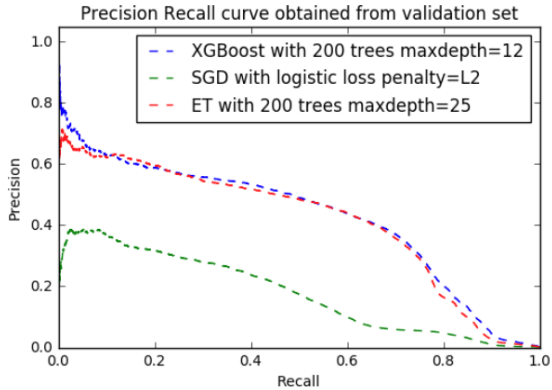

Figure 2: Precision recall curves for our models

## 3. CONCLUSION

In this paper, we describe our approach to Wiki Vandalism detection. We filter out invalid features and propose spam count features, which improve overall prediction performance. Our final solution utilizes state-of-the-art gradient boosting framework and demostrate the effctiveness of our method.

## 4. ADDITIONAL AUTHORS

## 5. REFERENCES

[1] L. Bottou. *Large-Scale Machine Learning with Stochastic Gradient Descent*, pages 177–186. Physica-Verlag HD, Heidelberg, 2010.
[2] T. Chen and C. Guestrin. Xgboost: A scalable tree boosting system. In *SIGKDD*, KDD '16, pages 785–794, New York, NY, USA, 2016. ACM.
[3] P. Geurts, D. Ernst, and L. Wehenkel. Extremely randomized trees. *Mach. Learn.*, 63(1):3–42, Apr. 2006.
[4] S. Heindorf, M. Potthast, H. Bast, B. Buchhold, and E. Haussmann. WSDM Cup 2017: Vandalism Detection and Triple Scoring. In *WSDM*. ACM, 2017.
[5] S. Heindorf, M. Potthast, B. Stein, and G. Engels. Vandalism detection in wikidata. In *CIKM*, pages 327–336. ACM, 2016.
[6] F. Pedregosa, G. Varoquaux, A. Gramfort, V. Michel, B. Thirion, O. Grisel, M. Blondel, P. Prettenhofer, R. Weiss, V. Dubourg, J. Vanderplas, A. Passos, D. Cournapeau, M. Brucher, M. Perrot, and E. Duchesnay. Scikit-learn: Machine learning in python. *J. Mach. Learn. Res.*, 12:2825–2830, Nov. 2011.